\documentclass[
english,
pra,
twocolumn,
superscriptaddress,
showpacs,
showkeys,
floatfix]
{revtex4-1}
\usepackage{graphicx}
\usepackage{bm}
\usepackage{textcomp}
\usepackage[usenames]{color}
\usepackage{amsfonts}
\usepackage{amssymb}
\usepackage{amsthm}
\usepackage{amsmath}
\usepackage{fontenc}
\usepackage[latin1]{inputenc}
\usepackage{latexsym}
\usepackage{times}

\newcommand{\ketb}[1]{|\kern.3ex#1\kern.3ex\rangle}
\newcommand{\brab}[1]{\langle\kern.3ex #1 \kern.3ex|}
\newcommand{\scalar}[2]{\langle\kern.3ex #1\kern.3ex|\kern.3ex#2\kern.3ex\rangle}

\renewcommand\Re{\operatorname{Re}}
\renewcommand\Im{\operatorname{Im}}

\DeclareMathAlphabet{\mathscr}{OT1}{pzc}{m}{it}

\begin{document}

\title{Coherent states and related quantizations for unbounded motions}

\author{V. G. Bagrov}
\email{bagrov@phys.tsu.ru}
\affiliation{Department of Physics, Tomsk State University, 634050, Tomsk, Russia.}
\altaffiliation[Also at ]{Institute of Physics, University of S\~ao Paulo, Brazil}
\author{J.-P. Gazeau}
\email{gazeau@apc.univ-paris7.fr}
\affiliation{Astroparticules et Cosmologie, Univ Paris Diderot, Sorbonne Paris Cit\'e, 75205 Paris, France}
\author{D. M. Gitman}
\email{gitman@dfn.if.usp.br}
\affiliation{Institute of Physics, University of S\~ao Paulo, Brazil}
\author{A. D. Levin}
\email{alevin@if.usp.br}
\affiliation{Institute of Physics, University of S\~ao Paulo, Brazil}
\begin{abstract}
We build coherent states (CS) for unbounded motions along two different
procedures. In the first one we adapt the Malkin-Manko construction for
quadratic Hamiltonians to the motion of a particle in a linear potential. A
generalization to arbitrary potentials is discussed. The second one extends
to continuous spectrum previous constructions of action-angle coherent
states in view of a consistent energy quantization.
\end{abstract}

\maketitle

\section{Introduction}

At present, coherent states (CS) take an important place in modern quantum
mechanics. They have a wide range of applications, in semiclassical
description of quantum systems, at the same time in quantization of
classical models, in condensed matter physics, in radiation theory, in loop
quantum gravity, and so on \cite{CSSI12}. In view of this wide range of
domains, a universally accepted definition of CS for arbitrary physical
systems and a universally accepted construction for them are still lacking.
Due to Glauber and Malkin-Manko (see \cite{Glauber,MalMa}) there exists a
well-defined construction algorithm for systems with quadratic Hamiltonians
(QH) with discrete spectra, and due to Gilmore, Perelomov and others (see
\cite{gilm72,perel72,perel86,AAG00}) for systems with a given Lie group
symmetry. Approaches based on action-angle formalism \cite{gaklau99} or on
reproducing kernel combined with Bayesian probabilistic ingredients \cite%
{agh08,gazeau09} have been developed more recently. In any generalization,
one attempts to maintain some of the basic properties of already known CS
for quadratic systems, like resolution of the unity. One of the most popular
constraints concerns semi-classical features. One thus attempts to maintain
saturation of uncertainty relations for some physical quantities (e.g.
coordinates and momenta) as they are given at a certain instant. One
requires that means of particle coordinates, calculated with respect to
time-dependent CS, move along the corresponding classical trajectories. In
addition, CS have to be labeled by parameters that have a direct classical
analog, let say by phase-space coordinates. It is also desirable for
time-dependent CS to maintain their form under the time evolution. One last
but not the least constraint in the construction is to give these special
states a status of quantizer \textit{\`a la} Berezin-Klauder \cite%
{ber75,klau95,klau00,gazeau09}.

As was already mentioned, usually CS are constructed for systems with
discrete energy spectra, which represent bounded motions: we thus pass from
quantum stationary states labelled with quantum numbers to quantum CS
labelled by phase space variables. There exist some attempts to construct CS
for systems with continuous spectra, see for instance \cite%
{hongoh77,gaklau99,gelklau09,GueLoA11}.
However, one can state that the problem is still open or at least deserves
to be examined in a more comprehensive way, particularly in view of
application to realistic systems. 

In this article, we examine the problem from two viewpoints. On one side we
adopt the approach of Malkin-Manko to systems with continuous spectra. On
the other side, we generalize, modify, and apply the approach followed in
\cite{gaklau99} to the same kind of systems. It should be noted that in the
first approach we start with a well-defined quantum formulation (canonical
quantization) of the physical system and the construction of coherent states
follows from such a quantization. In the second approach, the quantization
procedure is inherent to the CS construction itself. In both approaches we
pretend to construct CS for concrete systems with continuous spectra, free
one-dimensional particle, charged particle on the plane and submitted to an
electric field, and eventually one-dimensional particle submitted to an
arbitrary scattering potential.

\section{CS for QH systems with continuous spectra. A possible approach}

\subsection{An instructive example: a particle in a constant external force}

\subsubsection{Creation and annihilation operators-integrals of motion}

Let us consider the quantum motion of a particle subjected to a constant
force that is directed along the axis $x^{1}.$ In fact, it is enough to
consider only the one-dimensional motion in the $x^{1}$-direction, since the
motions in the $x^{2}$- and $x^{3}$-directions are separated and are free
motions. The quantum motion in the $x^{1}$-direction is described by the
one-dimensional Schrödinger equation of the form
\begin{equation}
i\hbar \frac{\partial \Psi }{\partial t}=\hat{H}\Psi ,\ \ \hat{H}=-\frac{%
\hbar ^{2}\partial _{x^{1}}^{2}}{2m}+\alpha x^{1},  \label{a1}
\end{equation}%
where the constant $\alpha $ determines the magnitude of the force.
Introducing dimensionless variables $x$ and$\,\tau $ as
\begin{equation}
x^{1}=lx,\ \tau =\frac{\hbar }{2ml^{2}}t,  \label{a2}
\end{equation}%
where $l$ is an arbitrary constant of the length dimension, Eq. (\ref{a2})
reduces to:
\begin{equation}
i\frac{\partial \Psi }{\partial \tau }=\hat{H}\Psi ,\ \ \hat{H}=\sqrt{2}%
bx-\partial _{x}^{2},\ b=\frac{\sqrt{2}ml^{3}\alpha }{\hbar ^{2}}.
\label{a3}
\end{equation}%
Hence we are left with one single dimensionless constant $b$.

We note that the classical trajectory of the position $x$ has the form%
\begin{equation}
x\left( \tau \right) =x_{0}+p_{0}\tau -\sqrt{2}b\tau ^{2},  \label{a4}
\end{equation}%
where $x_{0}$ and $p_{0}$ are arbitrary constants (initial data).

In spite of the fact that the Hamiltonian $\hat{H}$ has a continuous
spectrum, \textrm{spec}\ $\hat{H}=\mathbb{R},$ it is convenient to introduce
in the problem the familiar creation and annihilation operators $a^{\dag}$
and $a$ as follows:
\begin{gather}
a=\frac{x+\partial _{x}}{\sqrt{2}},\ \ a^{\dag}=\frac{x-\partial _{x}}{\sqrt{%
2}}\Rightarrow \nonumber\\
x=\frac{a+a^{\dag}}{\sqrt{2}},\ \ \partial _{x}=\frac{%
a-a^{\dag}}{\sqrt{2}};\ \ [a,a^{\dag}]=1.  \label{a6}
\end{gather}%
We recall well-known commutators of such operators which will be useful for
the sequel:
\begin{gather}
\lbrack a^{\dag}a,a] =[aa^{\dag},a]=-a,\nonumber\\
[a^{\dag}a,a^{\dag}]=[aa^{\dag},a^{\dag}]=a^{\dag},
\lbrack a^{n},a^{\dag}] = na^{n-1}, \nonumber\\
[(a^{\dag})^{n},a]=-n(a^{\dag})^{n-1}.  \label{a7}
\end{gather}%
When written in terms of the operators (\ref{a6}), the Hamiltonian takes the
form
\begin{equation}
\hat{H}=\frac{1}{2}(aa^{\dag}+a^{\dag}a-a^{2}-a^{\dag2})+b(a+a^{\dag}).
\label{a8}
\end{equation}%
The term $a^{2}-a^{\dag2}$ impedes the Hamiltonian to be reduced to an
oscillator-like form through a canonical transformation, which indicates
that there does not exist a ground state and the spectrum of $\hat{H}$ is
continuous.

For the oscillator-like quadratic Hamiltonians, CS are constructed with the
aid of a Fock discrete basis issued from the action of the creation
operators on the vacuum state $|0\rangle $ ($a|0\rangle =0$). Then the
Glauber-type instantaneous CS have the form $|z\rangle =D\left( z\right)
|0\rangle ,$ where the unitary operator $D\left( z\right) $ reads%
\begin{equation*}
D\left( z,a,a^{\dag }\right) =\exp \left\{ za^{\dag }+z^{\ast }a\right\} .
\end{equation*}%
In the course of the evolution the CS maintain their form with some time
dependent $z\left( t\right).$ The Malkin-Manko-type CS can be defined as
eigenvectors of some annihilation operators that are integrals of motion,
see \cite{MalMa}. In fact both constructions coincide for quadratic
Hamiltonians. In the case under consideration, it does not exist a
generalization of the Glauber construction, because of the absence of the
vacuum vector. However, the Malkin-Manko idea can be implemented, as we
describe below.

Let us construct an operator
\begin{equation}
\hat{A}\left( \tau \right) =f(\tau )a+g(\tau )a^{\dag }+\varphi (\tau ),
\label{a9}
\end{equation}%
where the functions $f(\tau ),\,g(\tau )$, and$\,\varphi (\tau )$ have to be
determined by demanding that the operator $\hat{A}\left( \tau \right) $ be
integral of motion of the equation (\ref{a3}). To this end operator $\hat{A}$
has to obey the condition%
\begin{equation}
\lbrack \hat{S},\hat{A}\left( \tau \right) ]=0,\ \ \hat{S}=i\frac{\partial }{%
\partial \tau }-\hat{H}.  \label{a10}
\end{equation}%
Using relations (\ref{a7}), one can see that the conditions (\ref{a10})
holds if the functions $f(\tau ),\,g(\tau ),\,\varphi (\tau )$ are solutions
to the system
\begin{equation}
i\dot{f}+f+g=0,\ \ i\dot{g}-f-g=0,\ \ i\dot{\varphi}+b(f-g)=0.  \label{a11}
\end{equation}%
The general solution of eqs. (\ref{a11}) has the form
\begin{gather}
f(\tau )=c_{1}+i(c_{1}+c_{2})\tau , \ \
g(\tau )=c_{2}-i(c_{1}+c_{2})\tau ,
\notag \\
\varphi (\tau )=b\tau \lbrack i(c_{1}-c_{2})-(c_{1}+c_{2})\tau ]+c_{3}
\notag \\
\ =b\tau \left\{ \left[ f(\tau )+g(\tau )\right] \tau +i\left[ f(\tau
)-g(\tau )\right] \right\} +c_{3},  \label{a12}
\end{gather}%
where $c_{j},\ j=1,\,2,\,3,$ are arbitrary complex constants. Without loss
of generality, we can set $c_{3}=0$.

We note that there is no nontrivial solution to (\ref{a11})\ that satisfies
the condition $f(\tau )=g(\tau )$.

It follows from Eqs. (\ref{a6}) and (\ref{a9}) that
\begin{equation}
\lbrack \hat{A}\left( \tau \right) ,\hat{A}^{\dag}\left( \tau \right)
]=\Delta  =|f(\tau )|^{2}-|g(\tau )|^{2}
=|c_{1}|^{2}-|c_{2}|^{2}.  \label{a13}
\end{equation}

If $\Delta >0$, then, without loss of generality, we can set $\Delta =1,$
which corresponds to the multiplication of $\hat{A}$ by a complex number. In
this case the operators $\hat{A}^{\dag}\left( \tau \right) $ and $\hat{A}%
\left( \tau \right) $ are familiar creation and annihilation operators.

If $\Delta =0$, then, without loss of generality, the operator $\hat{A}%
\left( \tau \right) $ can be considered as a self-adjoint one. ($\hat{A}%
\left( \tau \right) $ can differ from a self-adjoint one only by a complex
factor only). In this case, Eqs. (\ref{a12}) contain only one complex
constant $c$ and have the form%
\begin{gather}
f(\tau )=c+i(c+c^{\ast })\tau ,\ \
g(\tau )=f^{\ast }(\tau ),  \notag \\
\varphi (\tau )=b[i(c-c^{\ast })\tau -(c+c^{\ast })\tau ^{2}], \ \
\varphi (\tau )=\varphi ^{\ast }(\tau ).  \label{a14}
\end{gather}

Finally, if $\Delta <0$, then one has to treat $\hat{B}=\hat{A}^{\dag}$ as
an annihilation operator and we again have the case $\Delta >0$. Therefore,
in fact, we have to study only two cases: $\Delta =1,\,\Delta =0$.

\subsubsection{Coherent states}

Let us consider solutions $\psi (\tau ;x)$ of the equation (\ref{a3}) that,
at the same time, are eigenstates of the operator $\hat{A}\left( \tau
\right) ,$ with the eigenvalues $Z,$
\begin{equation}
\hat{A}\left( \tau \right) \psi (\tau ;x)=Z\psi (\tau ;x).  \label{a15}
\end{equation}

Let us consider the case $\Delta =1$. Here, we have a family of operators $%
\hat{A}\left( \tau \right) =\hat{A}\left( \tau ,c_{1},c_{2}\right) $
parametrized by complex numbers $c_{1}$ and $c_{2}$ such that $%
|c_{1}|^{2}-|c_{2}|^{2}=1.$ One can see that the spectrum of any $\hat{A}%
\left( \tau \right) $ is continuous, \textrm{spec} $\hat{A}\left( \tau
\right) =\mathbb{C},$ and the eigenstate $\psi _{Z}^{c_{1},c_{2}}(\tau ;x)$
corresponding to $Z$ can be constructed in two ways.

The states $\psi _{Z}^{c_{1},c_{2}}(\tau ;x)$ can be simply found as
solutions of the differential equation (\ref{a15}), taking the operator $%
\hat{A}\left( \tau \right) $ in the coordinate representation (\ref{a9})
with account taken of (\ref{a6}) and (\ref{a12}). As a result we obtain:%
\begin{gather}
\psi _{Z}^{c_{1},c_{2}}(\tau ;x)=\frac{\exp R}{\sqrt{(f-g)\sqrt{\pi }}},\ \notag\\
R=\frac{f+g}{2(f-g)}\left( x+2b\tau ^{2}-\frac{\sqrt{2}Z}{f+g}\right)^2 \notag\\
\ +\frac{Z\left[ (f+g)Z-(f^{\ast }+g^{\ast })Z^{\ast }\right] }{2(f^{\ast
}+g^{\ast })}
-ib\tau \left( \sqrt{2}x+\frac{2b\tau ^{2}}{3}\right) .
\label{a17}
\end{gather}%
One can see that
\begin{equation*}
\Re{R} =-\frac{q^{2}}{2},\ q=\frac{x-x(\tau )}{|f-g|}\,,
\end{equation*}
and
\begin{equation}  \label{classtraj}
x(\tau)=\frac{1}{\sqrt{2}}\left[ Z(f^{\ast }-g^{\ast })+Z^{\ast
}(f-g)-2b\tau ^{2}\right].
\end{equation}

The function $x(\tau )$ is just the classical trajectory (\ref{a4}) with the
initial data%
\begin{align}
&&x_{0} =\frac{1}{\sqrt{2}}\left[ (c_{1}-c_{2})Z^{\ast }+(c_{1}^{\ast
}-c_{2}^{\ast })Z\right] ,  \notag \\
&&p_{0} =i\sqrt{2}\left[ (c_{1}+c_{2})Z^{\ast }-(c_{1}^{\ast }+c_{2}^{\ast
})Z\right] .  \label{a18a}
\end{align}

For fixed complex numbers $c_{1}$ and $c_{2},$ under the condition $\Delta
=1,$ there is an one-to-one correspondence between the complex number $Z$
and the initial data $x_{0}$ and $p_{0},$
\begin{equation}
Z=\frac{c_{1}+c_{2}}{\sqrt{2}}x_{0}+\frac{i(c_{1}-c_{2})}{2\sqrt{2}}p_{0}\ .
\label{19b}
\end{equation}
The second way to construct the states $\psi _{Z}^{c_{1},c_{2}}(\tau ;x)$ is
reminiscent of the Glauber construction of CS. We define the vacuum state $%
|0,\tau \rangle $ for the operator $\hat{A}\left( \tau \right) ,$%
\begin{equation}
\hat{A}\left( \tau \right) |0,\tau \rangle =0,  \label{19}
\end{equation}%
and the unitary displacement operator $D\left( Z,\tau \right) ,$%
\begin{equation}
D\left( Z,\tau \right) =\exp \{Z\hat{A}^{\dag }\left( \tau \right) -Z^{\ast }%
\hat{A}\left( \tau \right) \}.  \label{20}
\end{equation}%
Then, the states (\ref{a17}) can be represented as%
\begin{equation}
\psi _{Z}(\tau ;x)=|Z,\tau \rangle =D\left( Z,\tau \right) |0,\tau \rangle .
\label{21}
\end{equation}%
We will call the states (\ref{a17}) or (\ref{21}) the coherent states (CS)
in the case under consideration.

Let us fix complex numbers $c_{1}$ and $c_{2}.$ Then, the CS (\ref{a17}) are
square integrable and normalized to the unity,%
\begin{equation}
\langle Z,\tau |Z,\tau \rangle =1.  \label{b1}
\end{equation}%
But they are not orthogonal, their overlapping relation has the form%
\begin{gather}
\langle Z^{\prime },\tau |Z,\tau \rangle =
\int_{-\infty }^{\infty }\left[\psi _{Z^{\prime }}^{c_{1},\,c_{2}}(\tau ;x)\right] ^{\ast }\psi_{Z}^{c_{1},\,c_{2}}(\tau ;x)dx=\notag \\
\exp \left( F/2\right) ,  \ \
F=Z\left( Z^{\prime \ast }-Z^{\ast }\right) +Z^{\prime \ast }\left(
Z-Z^{\prime }\right) .  \label{b3}
\end{gather}%
At any fixed $c_{1}$ and $c_{2}$ the CS for an overcomplete system with the
following resolution of the unity%
\begin{gather}
\int \frac{d^{2}Z}{\pi }\,\left[ \psi _{Z}^{c_{1},\,c_{2}}(\tau ;x^{\prime })%
\right] ^{\ast }\psi _{Z}^{c_{1},c_{2}}(\tau ;x)=
\delta (x-x^{\prime }), \ \ \notag \\
d^{2}Z=d\Re{Z}\,d\mathrm{\Im{Z}}. \nonumber
\end{gather}

\begin{figure}[!]
\begin{center}
\includegraphics[width=10cm]{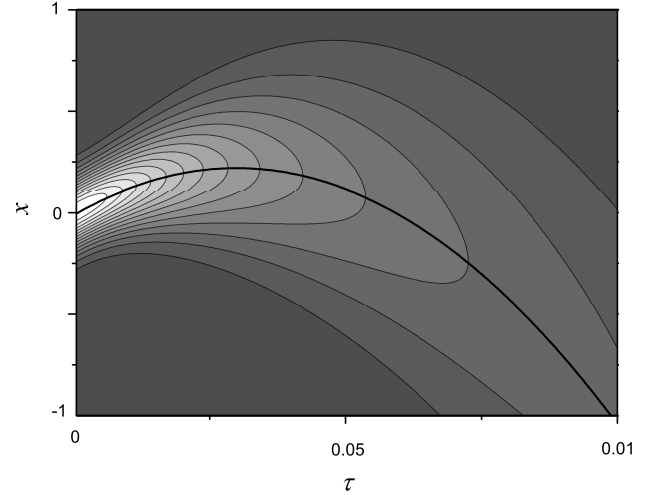}
\end{center}
\caption{Function $|\protect\psi _{Z}^{c_{1},c_{2}}(\protect\tau %
;x)|^{2},\;c_{1}=3,\;c_{2}=\sqrt{8},\;x_{0}=0,\;p_{0}=15,\;b=180.$ Plain line
represents the parabolic classical trajectory $x(\protect\tau)$ given by (%
\protect\ref{classtraj}).}
\label{fig3giba}
\end{figure}

To give some insight into the shape of these states and the way their
spreading faithfully follows the classical trajectory (\ref{classtraj}), we
show in Figure \ref{fig3giba} the time $\tau$ evolution of the probability
distribution $x\mapsto \vert \psi _{Z}^{c_{1},c_{2}}(\tau ;x)\vert^2$ for
some fixed values of other parameters.

\subsubsection{Semi-classical features}

Let us calculate some means and dispersions in the CS. To this end, we use
relations between the operators $\hat{x}$ and $\hat{p}=-i\partial _{x}$, and
the creation and annihilation operators $\hat{A}^{\dag }\left( \tau \right) $
and $\hat{A}\left( \tau \right) ,$ which follow from (\ref{a6}) and (\ref{a9}%
),%
\begin{gather}
\hat{x} =\frac{1}{\sqrt{2}}\left[ (f-g)(\hat{A}^{\dag}-\varphi ^{\ast})+(f^{\ast }-g^{\ast })(\hat{A}-\varphi )\right] ,  \notag \\
\hat{p} =i\frac{1}{\sqrt{2}}\left[ (f+g)(\hat{A}^{+}-\varphi ^{\ast})-(f^{\ast }+g^{\ast })(\hat{A}-\varphi )\right] .  \label{a31}
\end{gather}
Then%
\begin{gather}
\overline{x} \overset{\mbox{def}}{=}\langle Z,\tau |\hat{x}|Z,\tau \rangle
=x(\tau ) \notag \\
=\frac{1}{\sqrt{2}}\left[ Z(f^{\ast }-g^{\ast })+Z^{\ast }(f-g)-2b\tau ^{2}%
\right] ,  \notag \\
\overline{p} \overset{\mbox{def}}{=}\langle Z,\tau |\hat{p}|Z,\tau \rangle
=p(\tau )=\frac{p_{0}}{2}-\sqrt{2}b\tau .  \label{a32}
\end{gather}%
Let us introduce the deviation operators $\Delta x$ and $\Delta p$,
\begin{gather}
\Delta x =\hat{x}-x(\tau )= \notag \\
\frac{1}{\sqrt{2}}
\left[ (f-g)\left( \hat{A}^{+}-Z^{\ast }\right) +(f^{\ast }-g^{\ast })\left( \hat{A}-Z\right) \right], \notag \\
\Delta p =\hat{p}-p(\tau )=\notag \\ \frac{i}{\sqrt{2}}
\left[ (f+g)\left( \hat{A}^{+}-Z^{\ast }\right) -(f^{\ast }+g^{\ast })\left( \hat{A}-Z\right) \right] , \label{a33}
\end{gather}%
and the variances
\begin{gather}
\sigma _{1}=\overline{(\Delta x)^{2}},\ \ \sigma _{2}=\overline{(\Delta
p)^{2}}, \ \
\sigma_{3}=\frac{1}{2}\overline{(\Delta x\Delta p+\Delta p\Delta x)}.  \label{a36}
\end{gather}%
The latter quantities can be easily calculated:%
\begin{gather}
\sigma _{1}=\frac{1}{2}|f-g|^{2},\ \ \sigma _{2}=\frac{1}{2}|f+g|^{2}, \ \
\sigma _{3}=\frac{i}{2}(gf^{\ast }-g^{\ast }f).  \label{a37}
\end{gather}%
One can see that the variances do not depend on $Z,$ but depend on the
complex numbers $c_{1}$ and $c_{2}$ according to (\ref{a12}). Choosing the
numbers $c_{1}$ and $c_{2}$ one can provide any given (at $\tau =0$) value
for $\sigma _{1}$ or $\sigma _{2}$. It follows from (\ref{a37}):
\begin{equation}
J=\sigma _{1}\sigma _{2}-\sigma _{3}^{2}=1/4.  \label{a0}
\end{equation}%
The quantity $J$ does not depend on time, it is minimal in the CS.

One ought to mention that for $\alpha =0$ which correspond to the free
particle case, the CS (\ref{a17}) coincide with the ones constructed in the
work \cite{GueLoA11}.

\subsection{CS for general potentials}

Let the one-dimensional Schrödinger equation have a more general than (\ref%
{a1}) form
\begin{equation}
i\hbar \frac{\partial \Psi }{\partial t}=\hat{H}\Psi ,\ \ \hat{H}=-\frac{%
\hbar ^{2}\partial _{x^{1}}^{2}}{2m}+V\left( x^{1}\right) ,  \label{a1a}
\end{equation}%
where $V\left( x^{1}\right) $ is a potential which corresponds either to a
discrete or to a continuous spectrum. Using dimensionless variables $x$ and$%
\,\tau $ given by (\ref{a2}), we obtain
\begin{gather}
i\frac{\partial \Psi }{\partial \tau }=\hat{H}\Psi ,\ \ \hat{H}=-\partial
_{x}^{2}+U\left( x\right) ,\ \ \
 U\left( x\right) =\frac{2ml^{2}}{\hbar ^{2}}%
V\left( lx\right) .  \label{a2a}
\end{gather}

Let us suppose that we are able to construct an operator $\hat{A}\left( \tau
\right) $-integral of motion that obeys the conditions%
\begin{gather}
\lbrack \hat{S},\hat{A}\left( \tau \right) ]=0,\ \ \hat{S}=i\frac{\partial }{%
\partial \tau }-\hat{H},\ \ \
[\hat{A}\left( \tau \right) ,\hat{A}^{+}\left(
\tau \right) ]=1.  \label{a3a}
\end{gather}%
Then, we define the vacuum state $|0,\tau \rangle $ for the operator $\hat{A}%
\left( \tau \right) $ by equation (\ref{19}), the unitary displacement
operator $D\left( Z,\tau \right) $ given by eq. (\ref{20}), and finally
coherent states $D\left( Z,\tau \right) $ by equation (\ref{21}).

We stress that such a construction is based on the possibility to find a
complete discrete set of solutions of the Schrödinger equation with a given
potential. Such a possibility naively follows from the existence of a
unitary evolution operator in the case under consideration and from the
existence of a discrete complete basis in the corresponding Hilbert space
(then vectors from such a basis can be chosen as initial states and
developed then into a complete set of solutions by the evolution operator).
However, we know that the definition domain of the Hamiltonian as a rule
does not coincide with the Hilbert space, this is a source of numerous
paradoxes (see \cite{GVT}) and, in particular, can create difficulties with
the realization of the described above program.

For any quadratic potential $U\left( x\right) $ operators $\hat{A}\left(
\tau \right) $ and $\hat{A}^{+}\left( \tau \right) $ are expressed by a
linear canonical transformation with the creation and annihilation operators
$a^{\dag }$ and $a$ given by (\ref{a6}). Coefficient functions in such a
canonical transformation obey ordinary differential equations of second
order, \cite{BagGi90}. For more general potentials one has to elaborate
specific methods for solving the operator equations (\ref{a3a}). In any
case, in the approach under consideration, we are not restricted by the
demand that the system has to have a discrete spectrum.

\section{CS for conservative systems with continuous spectra. An alternative
construction}

\label{PAACS}

\subsection{Pseudo-action\,\&\,angle variables}

We consider again the motion of a particle of mass $m$ on the line, with
phase space conjugate variables $(q,p)$, and submitted to a potential $V(q)$%
. Suppose it conservative. For a given unbounded motion its Hamiltonian
function is fixed to a certain value $E$ of the energy:
\begin{equation}
H(q,p)=\frac{p^{2}}{2m}+V(q)=E\,.
\end{equation}%
Solving this for the momentum variable $p$, assuming a positive velocity,
leads to
\begin{equation}
p=p(q,E)=\sqrt{2m}\sqrt{E-V(q)}\,,
\end{equation}%
supposing no restriction on $q$, e.g. $E-V(q)>0$ for all $q$. From $p=mdq/dt$
we derive the expression of the time as a function of $(q,p)$, through $V$
and from $E=E(q,p)$:
\begin{gather}
dt=\sqrt{\frac{m}{2}}\frac{dq}{\sqrt{E-V(q)}}\quad \Rightarrow \notag \\
t-t_{0}=\sqrt{%
\frac{m}{2}}\int_{q_{0}}^{q}\frac{dq^{\prime }}{\sqrt{E-V(q^{\prime })}}\,.
\end{gather}%
We then introduce a \textquotedblleft pseudo-action\textquotedblright\
variable, depending on $(q,p)$ through the energy only, $\mathfrak{J}=%
\mathfrak{J}(E)$, with derivative submitted to the condition
\begin{equation}
\mathfrak{J}^{\prime }(E)=\frac{d\mathfrak{J}}{dE}>0\,.
\end{equation}%
Thus the map $E\mapsto \mathfrak{J}(E)$ is one-to-one and $E$ can be
considered as well as a function of $\mathfrak{J}$: $E=E(\mathfrak{J})$. We
now consider the map $(q,p)\mapsto (\mathfrak{J},t)$ with Jacobian matrix
\begin{equation}
\label{s.long}
\begin{pmatrix}
\frac{m}{p}-\frac{V^{\prime }(q)}{2}\sqrt{\frac{m}{2}}\int_{q_{0}}^{q}\frac{%
dq^{\prime }}{(E-V(q^{\prime 3/2}} & -\frac{p}{2}\sqrt{\frac{1}{2m}}%
\int_{q_{0}}^{q}\frac{dq^{\prime }}{(E-V(q^{\prime 3/2}} \\
\mathfrak{J}^{\prime }(E)\,V^{\prime }(q) & \mathfrak{J}^{\prime }(E)\,\frac{%
p}{m}%
\end{pmatrix}%
\,,
\end{equation}%
with determinant equal to $\mathfrak{J}^{\prime }(E)\equiv (F(\mathfrak{J}%
))^{-1}$. Therefore, the map
\begin{equation}  \label{qpjgam}
(q,p)\mapsto (\mathfrak{J},\gamma )\,,\quad \gamma \overset{\mbox{def}}{=}F(%
\mathfrak{J}(E(q,p))\,t(q,p)\,,
\end{equation}%
has Jacobian equal to 1, i.e. is canonical. New variables will be called
\textquotedblleft pseudo-action--angle\textquotedblright\ variables by
analogy with the usual action-angle variable used for bounded
one-dimensional motions. Note the role played by $\gamma$ as a kind of
intrinsic time for the system, like the angle variable does for bounded
motions.

Suppose that measurements on the considered one-dimensional system with
classical energy $E=p^{2}/2m+V(q)$ yield the continuous spectral values for
the energy observable (up to a constant shift), denoted by $\mathscr{E}$:
\begin{gather}  \label{enspec}
0\leq \mathscr{E}<\mathscr{E}_{M}\,,\quad \mathscr{E}_{M}\
\mbox{finite or
$\infty$}
\end{gather}%
The difference between the two physical quantities, classical $E$ and
quantum $\mathscr{E}$, lies in the probabilistic nature of the measurement
of the latter, involving Hilbertian quantum states. Let $\varepsilon $ be a
constant characteristic energy of the considered system (e.g. $h/\tau $,
where $\tau $ is a characteristic time). We put $\tilde{\mathscr{E}}=%
\mathscr{E}/\varepsilon $. We define a corresponding sequence of probability
distributions $\mathfrak{J}\mapsto p_{\mathscr{E}}(\mathfrak{J})$, $\int_{%
\mathcal{R}_{\mathfrak{J}}}d\tilde{\mathfrak{J}}\,p_{\mathscr{E}}(\mathfrak{J%
})=1$, supposing a (prior) \emph{uniform} distribution on the range $%
\mathcal{R}_{\mathfrak{J}}$ of the pseudo-action variable $\mathfrak{J}$.
Furthermore, we impose $p_{\mathscr{E}}(\mathfrak{J})$ to obey the two
conditions:%
\begin{gather}
0<\,\mathcal{N}(\mathfrak{J})\overset{\mbox{def}}{=}\int_{0}^{\tilde{%
\mathscr{E}}_{M}}d\tilde{\mathscr{E}}\,
p_{\mathscr{E}}(\mathfrak{J})<\infty
\,,\notag\\ \mathscr{E}=\int_{\mathcal{R}_{\mathfrak{J}}}d\tilde{\mathfrak{J}}%
\,E(\mathfrak{J})\,p_{\mathscr{E}}(\mathfrak{J})\,,
\end{gather}%
where $\tilde{\mathfrak{J}}=\mathfrak{J}/h$, $h$ being the Plank constant.
The finiteness condition allows to consider the map $\mathscr{E}\mapsto p_{%
\mathscr{E}}(\mathfrak{J})/\mathcal{N}(\mathfrak{J})$ as a probabilistic
model referring to the continuous energy data, which might viewed in the
present context as a \emph{prior distribution}.

\subsection{Pseudo-action-angle coherent states}

Let $\mathcal{H}$ be a complex Hilbert space with distributional orthonormal
basis $\{| \psi_{\mathscr{E}}\rangle\,, \, 0\leq \mathscr{E} < \mathscr{E}_M
\}$,
\begin{gather}  \label{smeas}
\langle \psi_{\mathscr{E}} | \psi_{{\mathscr{E}}^{\prime }}\rangle = \delta(%
\mathscr{E}-\mathscr{E}^{\prime })\, , \ \
 \int_0^{\tilde{\mathscr{E}}_M}d%
\tilde{\mathscr{E}}\, | \psi_{\mathscr{E}}\rangle\langle \psi_{\mathscr{E}}|
= 1_{\mathcal{H}}\, .
\end{gather}
The pseudo-action-angle phase space for the unbounded motion with measured
energies $0 \leq \mathscr{E}< \mathscr{E}_M$ is the set $X = \{(\mathfrak{J}%
,\gamma)\, , \, \mathfrak{J} \in \mathcal{R}_{\mathfrak{J}}\,, \, \gamma\in
\mathbb{R}\}$. Let $\left( p_{\mathscr{E}}(\mathfrak{J})\right)$ be the
continuous set of probability distributions associated with these energies.
One then constructs the family of states in $\mathcal{H}$ for the considered
motion as the following continuous map from $X$ into $\mathcal{H}$:
\begin{gather}
X \ni (\mathfrak{J},\gamma) \mapsto |\mathfrak{J},\gamma\rangle = \notag\\
\frac{1}{\sqrt{\mathcal{N}(\mathfrak{J})}}\int_0^{\tilde{\mathscr{E}}_M}d\tilde{%
\mathscr{E}}\, \sqrt{p_{\mathscr{E}}(\mathfrak{J})}\, e^{-i\alpha_{%
\mathscr{E}}\, \gamma}\, |\psi_{\mathscr{E}}\rangle \in \mathcal{H}\,,
\end{gather}
where the choice of the real function $\mathscr{E} \mapsto\alpha_{\mathscr{E}%
}$ is left to us in order to comply with some reasonable physical criteria.
A natural choice which guaranties time evolution stability is $\alpha_{%
\mathscr{E}} =\varsigma \tilde{\mathscr{E}}$, where $\varsigma$ is some
constant.

The coherent states $|\mathfrak{J},\gamma\rangle$ are unit vector : $\langle
\mathfrak{J},\gamma| J,\gamma\rangle = 1$ and resolve the unity operator in $%
\mathcal{H}$ with respect to the measure ``in the Bohr sense'' $\mu_B(d%
\mathfrak{J}\,d\gamma)$ on the phase space $X$ :
\begin{gather}
\int_{X} \mu_B(d\mathfrak{J}\,d\gamma)\, \mathcal{N}(\mathfrak{J})\,|
\mathfrak{J},\gamma\rangle\langle \mathfrak{J},\gamma|\overset{\mbox{def}}{=} \notag\\
\int_{\mathcal{R}_{\mathfrak{J}}} d\tilde{\mathfrak{J}}\, \mathcal{N}(%
\mathfrak{J})\,\lim_{T \to\infty}\frac{1}{T}\int_{-\frac{T}{2}}^{\frac{T}{2}%
} d\gamma | \mathfrak{J},\gamma\rangle\langle \mathfrak{J},\gamma| =1_{%
\mathcal{H}}\, .
\end{gather}
This property allows a \emph{coherent state quantization} of classical
observables $f(\mathfrak{J},\gamma)$ which is energy compatible with our
construction of the posterior distribution $\mathfrak{J} \mapsto p_{%
\mathscr{E}}(\mathfrak{J})$ in the following sense:
\begin{gather}
f(\mathfrak{J},\gamma) \mapsto
\int_{X} \mu_B(dJ\,d\gamma)\, \mathcal{N}(%
\mathfrak{J})\, f(\mathfrak{J},\gamma)\, | \mathfrak{J},\gamma\rangle\langle
\mathfrak{J},\gamma|\overset{\mbox{def}}{=} A_f \,
\end{gather}
Indeed, it is trivially verified that the quantum Hamiltonian is what we
expect:
\begin{gather*}
A_{H} = \int_{X} \mu_B(dJ\,d\gamma)\, \mathcal{N}(\mathfrak{J})\, E(%
\mathfrak{J})\, | \mathfrak{J},\gamma\rangle\langle \mathfrak{J}%
,\gamma|= \notag\\
\int_0^{\tilde{\mathscr{E}}_M}d\tilde{\mathscr{E}}\, \mathscr{E}
|\psi_{\mathscr{E}} \rangle\langle \psi_{\mathscr{E}}|\,,
\end{gather*}
that is, the states $|\psi_{\mathscr{E}} \rangle$ are eigendistributions of
the quantum Hamiltonian $A_{H}$ with eigenvalues the elements of the
spectrum (\ref{enspec}).

The quantization of any function $f(\mathfrak{J})$ of the single
pseudo-action variable yields the diagonal operator:
\begin{equation}  \label{ }
f(\mathfrak{J}) \mapsto A_f = \int_0^{\tilde{\mathscr{E}}_M}d\tilde{%
\mathscr{E}}\, \langle f\rangle_\mathscr{E} |\psi_{\mathscr{E}}
\rangle\langle \psi_{\mathscr{E}}|\,.
\end{equation}
where
\begin{equation}
\langle f\rangle_{\mathscr{E}}= \int_{\mathcal{R}_{\mathfrak{J}}} d\tilde{%
\mathfrak{J}} \, f(\mathfrak{J}) \, p_{\mathscr{E}}(\mathfrak{J})\, .
\end{equation}
Alternatively, the quantization of any function $f(\gamma)$ of the single
angle variable only yields the operator:
\begin{gather}  \label{ }
f(\gamma) \mapsto A_f =
\int_0^{\tilde{\mathscr{E}}_M}d\tilde{\mathscr{E}}%
\int_0^{\tilde{E}^{\prime }_M}d\tilde{\mathscr{E}}^{\prime }\, [A_f]_{%
\mathscr{E} \mathscr{E}^{\prime }} |\psi_{\mathscr{E}} \rangle\langle \psi_{%
\mathscr{E}^{\prime }}|\, ,
\end{gather}
where the matrix elements$[A_f]_{\mathscr{E} \mathscr{E}^{\prime }} $ are
formally given by:
\begin{gather}
[A_f]_{\mathscr{E} \mathscr{E}^{\prime }} = \notag\\ \int_{\mathcal{R}_{\mathfrak{J}%
}} d\tilde{\mathfrak{J}} \, \sqrt{p_{\mathscr{E}}(\mathfrak{J})\,p_{%
\mathscr{E}^{\prime }}(\mathfrak{J})}\,\lim_{T \to\infty}\frac{1}{T}\int_{-%
\frac{T}{2}}^{\frac{T}{2}} d\gamma \, e^{-i(\alpha_{\mathscr{E}} - \alpha_{%
\mathscr{E}^{\prime }})\gamma} \, f(\gamma)\, .
\end{gather}
In particular the CS quantization procedure provides, for a given choice of
the function $\mathscr{E} \mapsto \alpha_{\mathscr{E}}$, a self-adjoint
operator corresponding to any real bounded or semi-bounded function $%
f(\gamma)$. For instance, the quantization of the elementary Fourier
exponential $f(\gamma) = e^{i\varpi \gamma}$ gives a bounded operator with
matrix elements (in the considered energy range):
\begin{gather}
[A_{e^{i\varpi \gamma}}]_{\mathscr{E} \mathscr{E}^{\prime }} =
\pi
\left\lbrack\int_{\mathcal{R}_{\mathfrak{J}}} d\tilde{\mathfrak{J}} \, \sqrt{%
p_{\mathscr{E}}(\mathfrak{J})\,p_{\mathscr{E}^{\prime }}(\mathfrak{J})}%
\right\rbrack\, \delta(\alpha_{\mathscr{E}^{\prime }} - \alpha_{\mathscr{E}}
+ \varpi)\, .
\end{gather}
The quantization of the original canonical position and momentum variables $%
(q,p)$ is carried out through the functions $q= q(\mathfrak{J}, \gamma)$, $%
p= p(\mathfrak{J}, \gamma)$ obtained through the inverse of the map (\ref%
{qpjgam}). It yields symmetric position and momentum operators.
Self-adjointness is not guaranteed, depending or not on the choice of the
choice of distribution $\mathfrak{J} \mapsto p_{\mathscr{E}}(\mathfrak{J})$
and the function $\mathscr{E} \mapsto \alpha_{\mathscr{E}}$. It is possible
that regularization techniques are needed here.

Semi-classical aspects of such coherent states and related quantization are
suitably caught through the so-called lower symbols of operators $A_f$, i.e.
their mean values in coherent states $\check{f}(\mathfrak{J},\gamma)=
\langle \mathfrak{J},\gamma| A_f|\mathfrak{J},\gamma\rangle$. As a matter of
fact, the map $f\mapsto \check{f}$ is the Berezin-like integral transform
\begin{gather}  \label{mapff}
\check{f}(\mathfrak{J},\gamma) =
\int_{X}
\mu_B(dJ^{\prime}\,d\gamma^{\prime})\, \mathcal{N}(\mathfrak{J}^{\prime})\,
f(\mathfrak{J}^{\prime},\gamma^{\prime})\,\vert \langle \mathfrak{J}^{\prime
},\gamma^{\prime }|\mathfrak{J} ,\gamma\rangle \vert^2\, ,
\end{gather}
which gives at once some insight on the domain properties of $A_f$ and on
the semi-classical behavior of the coherent states.

\subsection{An exploration with normal law}

Let us choose the following function for the classical pseudo-action:
\begin{equation}  \label{ }
\tilde{\mathfrak{J}} (E) = \eta \ln\tilde{E} \, , \ \Leftrightarrow \ \tilde{%
E} = e^{\tilde{\mathfrak{J}}/\eta}\, , \quad \eta>0\, ,\tilde{E} >0\, ,
\end{equation}
and so $\mathcal{R}_{\mathfrak{J}} = \mathbb{R}$ for the range of $\mathfrak{%
J}$. For the probability distribution $\mathfrak{J} \mapsto p_{\mathscr{E}}(%
\mathfrak{J})$ we choose the normal law centered at $\eta \ln\tilde{%
\mathscr{E}}$:
\begin{equation}
p_{\mathscr{E}}(\mathfrak{J}) = \left(\frac{\epsilon}{\pi}\right)^{1/2}\,
e^{-\epsilon (\tilde{\mathfrak{J}} - \eta \ln \tilde{\mathscr{E}})^2}\, .
\end{equation}
Then the three fundamental requirements are (almost) fulfilled:

\begin{itemize}
\item[(i)] it is probabilistic: $\int_{\mathbb{R}}d\tilde{\mathfrak{J}}\, p_{%
\mathscr{E}}(\mathfrak{J}) = 1$,

\item[(ii)] the average value of the classical energy is $\approx$ the
observed value at large $\epsilon$ or $\eta$:
\begin{equation}
\int_{\mathbb{R}}d\tilde{\mathfrak{J}}\, E(\mathfrak{J})\, p_{\mathscr{E}}(%
\mathfrak{J}) = e^{\frac{1}{4\epsilon \eta^2}} \, \mathscr{E} \, ,
\end{equation}

\item[(iii)] positiveness and finiteness conditions are fulfilled:
\begin{gather}
0< \,\mathcal{N} (\mathfrak{J}) =
 \int_{0}^{\tilde{\mathscr{E}}_M}d\tilde{%
\mathscr{E}} \, p_{\mathscr{E}}(\mathfrak{J}) = \frac{1}{\eta} \,e^{\left(%
\frac{\tilde{\mathfrak{J}}}{\eta} + \frac{1}{4\epsilon \eta^2}\right)} <
\infty\, .
\end{gather}
\end{itemize}

Note the average value of $\mathfrak{J}$ : $\int_{\mathbb{R}}d\tilde{%
\mathfrak{J}}\, \mathfrak{J}\, p_{\mathscr{E}}(\mathfrak{J})= h\eta \ln%
\tilde{\mathscr{E}}$. Coherent states with $\alpha_{\mathscr{E}}= \varsigma%
\tilde{\mathscr{E}}$ read as:
\begin{gather}  \label{CSNLA}
X= \mathbb{R}\times \mathbb{R} \mapsto \notag\\
|\mathfrak{J},\gamma\rangle =
\frac{1%
}{\sqrt{\mathcal{N}(\mathfrak{J})}}\int_0^{\infty}d\tilde{\mathscr{E}}\,
\sqrt{p_{\mathscr{E}}(\mathfrak{J})}\, e^{-i\varsigma\tilde{\mathscr{E}}\,
\gamma}\, |\psi_{\mathscr{E}}\rangle \in \mathcal{H}\,,
\end{gather}
They are, by construction, unit vectors, are temporal evolution stable for
large $\epsilon$ or $\eta$, and solve the identity:
\begin{gather}  \label{ }
\langle \mathfrak{J},\gamma |\mathfrak{J},\gamma\rangle = 1\, ,
 \quad
\int_{X} \mu_B(d\mathfrak{J}\,d\gamma)\, \mathcal{N}(\mathfrak{J})\,|
\mathfrak{J},\gamma\rangle\langle \mathfrak{J},\gamma|= 1_{\mathcal{H}}\, , \notag\\
\quad e^{-i\tilde{A}_Ht} |\mathfrak{J},\gamma\rangle = |\mathfrak{J},\gamma
+ t/\varsigma\rangle\, ,
\end{gather}
with $\tilde{A}_H = A_H/h$. They overlap as
\begin{gather}  \label{CSNLAov}
\langle \mathfrak{J}^{\prime },\gamma^{\prime }|\mathfrak{J},\gamma\rangle =
\frac{1}{\sqrt{\mathcal{N}(\mathfrak{J}) \mathcal{N}(\mathfrak{J}^{\prime })}%
}\, e^{-\frac{\epsilon}{4}(\tilde{\mathfrak{J}} - \tilde{\mathfrak{J}}%
^{\prime })^2}\, \left(\frac{\epsilon}{\pi}\right)^{1/2}\, \notag\\ \times \int_0^{+\infty}d%
\tilde{\mathscr{E}} \, e^{-i\varsigma\tilde{\mathscr{E}}(
\gamma-\gamma^{\prime })}\, e^{-\epsilon\left(\frac{\tilde{\mathfrak{J}} +
\tilde{\mathfrak{J}}^{\prime }}{2}-\eta \ln \tilde{\mathscr{E}}\right)^2}\, .
\end{gather}
This indicates a bell-shaped localization in pseudo-action variable at large
$\tilde{\mathfrak{J}}$ or at large $\epsilon$:
\begin{gather}  \label{CSNLAov1}
\vert \langle \mathfrak{J}^{\prime },\gamma^{\prime }|\mathfrak{J}%
,\gamma\rangle \vert\leq \frac{1}{\sqrt{\mathcal{N}(\mathfrak{J}) \mathcal{N}%
(\mathfrak{J}^{\prime })}}\, e^{-\frac{\epsilon}{4}(\tilde{\mathfrak{J}} -
\tilde{\mathfrak{J}}^{\prime })^2}\, \left(\frac{\epsilon}{\pi}%
\right)^{1/2}\, \notag\\
\times \int_0^{+\infty}d\tilde{\mathscr{E}} \, e^{-\epsilon\left(%
\frac{\tilde{\mathfrak{J}} + \tilde{\mathfrak{J}}^{\prime }}{2}-\eta \ln
\tilde{\mathscr{E}}\right)^2}\, .
\end{gather}
A similar good localization in angle requires a study of the behavior at
large $k$ of the following Fourier transform: $\int_0^{+\infty} dx
e^{-ikx}\, e^{-\mu(\ln x -\lambda)^2}\, , $ with $x= \tilde{\mathscr{E}}$, $%
k=\varsigma(\gamma-\gamma^{\prime })$, $\mu = \epsilon \eta^2$, and $\lambda=%
\frac{\tilde{\mathfrak{J}} + \tilde{\mathfrak{J}}^{\prime }}{2\eta}$.

An interesting observation concerns the CS quantization of any power of the
classical energy:
\begin{align*}
A_{H^{\lambda}} &= \int_{X} \mu_B(dJ\,d\gamma)\, \mathcal{N}(\mathfrak{J})\,
(E(\mathfrak{J}))^{\lambda}\, | \mathfrak{J},\gamma\rangle\langle \mathfrak{J%
},\gamma| \\
&=e^{\frac{\lambda^2}{4\epsilon \eta^2}}\int_0^{+\infty}d\tilde{\mathscr{E}}%
\, ( \mathscr{E})^{\lambda} |\psi_{\mathscr{E}} \rangle\langle \psi_{%
\mathscr{E}}|\, ,
\end{align*}
which means that $A_{H^{\lambda}} = e^{\frac{\lambda(\lambda -1)}{4\epsilon
\eta^2}}\left(A_H\right)^{\lambda}$. The quantization of the Fourier
exponential $e^{i \varpi \gamma}$ gives the bounded operator
\begin{gather}
A_{e^{i\varpi \gamma}}=
\frac{\pi}{\varsigma}\int_{\sup(0,
\varpi/\varsigma)}^{+\infty}e^{-\frac{\epsilon \eta}{4}\left(\ln\left(\frac{%
\tilde{\mathscr{E}}}{\tilde{\mathscr{E}} - \varpi/\varsigma}%
\right)\right)^2}\, |\psi_{\mathscr{E}} \rangle\langle \psi_{\mathscr{E}%
-\varpi/\varsigma}|\, .
\end{gather}
We might be able to deduce from this formula the quantization of the
variable $\gamma$ by the formal trick $A_{\gamma}= \left.-i
\partial/\partial\varpi A_{e^{i\varpi \gamma}}\right\vert_{\varpi=0}$.

\subsection{Probability distributions on phase or other spaces}

\begin{figure}[tbp]
\begin{center}
\includegraphics[width=9.5cm]{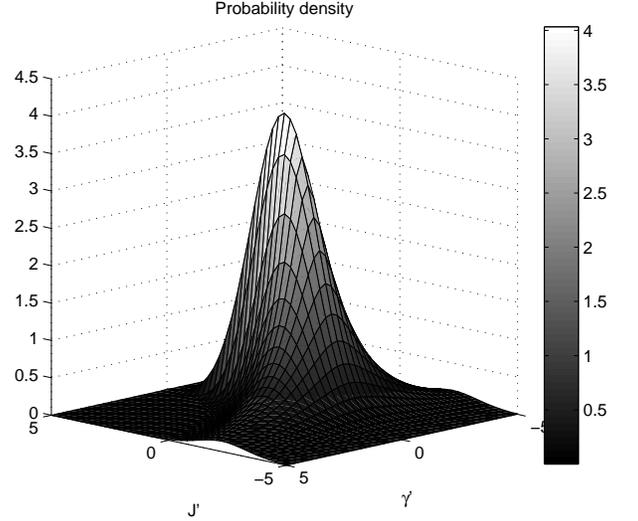}
\end{center}
\caption{Calculated phase space probability density (Eq. \protect\ref%
{probdistpsA}) with $\mathfrak{J}=\protect\gamma=0$, $\protect\varsigma=%
\protect\epsilon=\protect\eta=1$ and zoom in the range $\mathfrak{J}^{\prime
}: -5 \div 5\,,\quad \protect\gamma^{\prime }: -5 \div 5$.}
\label{probdensfig}
\end{figure}

In Figure \ref{probdensfig} are shown two-dimensional pictures of the
probability density $\mathcal{N}(\mathfrak{J})\vert \langle \mathfrak{J}%
^{\prime },\gamma^{\prime }|\mathfrak{J},\gamma\rangle\vert^2$ with
\begin{equation}
\mathcal{N} (\mathfrak{J}) = \int_{0}^{\tilde{\mathscr{E}}_M}d\tilde{%
\mathscr{E}} \, p_{\mathscr{E}}(\mathfrak{J}) = \frac{1}{\eta} \,e^{\left(%
\frac{\tilde{\mathfrak{J}}}{\eta} + \frac{1}{4\epsilon \eta^2}\right)}
\end{equation}
and,
\begin{gather}  \label{probdistpsA}
\langle \mathfrak{J}^{\prime },\gamma^{\prime }|\mathfrak{J},\gamma\rangle =
\frac{1}{\sqrt{\mathcal{N}(\mathfrak{J}) \mathcal{N}(\mathfrak{J}^{\prime })}%
}\, e^{-\frac{\epsilon}{4}(\tilde{\mathfrak{J}} - \tilde{\mathfrak{J}}%
^{\prime})^2}\, \left(\frac{\epsilon}{\pi}\right)^{1/2}\, \notag\\ \times \int_0^{+\infty}d%
\tilde{\mathscr{E}} \, e^{-i\varsigma\tilde{\mathscr{E}}(
\gamma-\gamma^{\prime })}\, e^{-\epsilon\left(\frac{\tilde{\mathfrak{J}} +
\tilde{\mathfrak{J}}^{\prime }}{2}-\eta \ln \tilde{\mathscr{E}}\right)^2}\, .
\end{gather}
We fix the parameters $(\mathfrak{J},\gamma)$ and sweep $(\mathfrak{J}%
^{\prime },\gamma^{\prime })$ in the $\mathbb{R}\times\mathbb{R}$ space.
Parameters $\epsilon=\eta=\varsigma=\mathfrak{J}=1$ are chosen for an
example. That gives a nice picture of the expected good localization of
these states in the phase space plane $(\mathfrak{J}^{\prime
},\gamma^{\prime })$.

Let us explore another representation, picking $\mathcal{H} = L^2(\mathbb{R})
$ as a companion Hilbert space and as continuous basis the
eigen-distributions of the operator $-d^2/d\mathsf{x}^2$ \cite{GVT}: to each
eigenvalue $\tilde{\mathscr{E}}$ correspond the symmetric $\psi_{\mathscr{E}%
}^+ (\mathsf{x})= \frac{1}{\sqrt{4\pi \sqrt{\tilde{\mathscr{E}}} }} \cos(%
\sqrt{\tilde{\mathscr{E}}} \mathsf{x})$ and the antisymmetric $\psi_{%
\mathscr{E}}^- (\mathsf{x})= \frac{-i}{\sqrt{4\pi \sqrt{\tilde{\mathscr{E}}}
}} \sin(\sqrt{\tilde{\mathscr{E}}} \mathsf{x})$ (the phase $-i$ is chosen
for convenience). A degeneracy of order 2 is present here and should be
taken into account by including a factor 2 in the spectral measure $d\tilde{%
\mathscr{E}}| \psi_{\mathscr{E}}\rangle\langle \psi_{\mathscr{E}}|$
appearing in (\ref{smeas}). We should caution against the risk of confusion
with the position representation: the symbol $\mathsf{x}$ should not be
regarded in general as an element of the spectrum of the position operator $%
A_q$, and instead, we should view the states (\ref{CSNLA}) as special wave
packets in representation ``$\mathsf{x}$''. We find from (\ref{CSNLA})
(after the change $u= \sqrt{\tilde{\mathscr{E}}}$),
\begin{gather}
\langle \mathsf{x} | \mathfrak{J}, \gamma\rangle = \sqrt{\frac{\epsilon\eta^2%
}{\pi^3}} \, e^{-\frac{\epsilon}{2}\left( \tilde{\mathfrak{J}} + \frac{1}{%
2\epsilon\eta}\right)^2} \notag\\ \times \int_0^{+ \infty} du\, u^{2\epsilon \eta \tilde{%
\mathfrak{J}} +1/2}\, e^{-2 \epsilon \eta^2 (\ln u)^2 }\, e^{-i(\varsigma
\gamma u^2 + u\mathsf{x})}\, ,
\end{gather}
The study of this expression amounts to analyze the behavior of the
following Fourier transform:
\begin{equation}
F( \mathsf{x})= \frac{1}{\sqrt{2\pi}} \int_0^{+\infty} du \, e^{-i \mathsf{x}%
u} \, u^{\alpha}\, e^{-\delta (\ln u)^2}\, e^{-i\beta u^2}\, ,
\end{equation}
with $\alpha = 2\epsilon \eta \tilde{\mathfrak{J}} + 1/2$, $\beta =
\varsigma \gamma$, and $\delta= 2\epsilon \eta^2 $. From the upper bound
\begin{equation}
\vert F( \mathsf{x})\vert \leq \sqrt{\frac{2}{\delta}}\, e^{\frac{%
(\alpha+1)^2}{4\delta}}\, ,
\end{equation}
we see that it can be made arbitrarily small at large $\eta$. The map
\begin{gather}  \label{locprobspat1}
\mathsf{x}\mapsto \vert \langle \mathsf{x} | \mathfrak{J}, \gamma\rangle
\vert^2 = \frac{\epsilon\eta^2}{\pi^3} \, e^{-\epsilon\left( \tilde{%
\mathfrak{J}} + \frac{1}{2\epsilon\eta}\right)^2} \notag \\ \times\left\vert  \int_0^{+
\infty} du\, u^{2\epsilon \eta \tilde{\mathfrak{J}} +1/2}\, e^{-2 \epsilon
\eta^2 (\ln u)^2 }\, e^{-i(\varsigma \gamma u^2 + \mathsf{x}u)}\right\vert^2
\end{gather}
defines a probability distribution on the real line. As shown in Figure \ref%
{probdensfigpos}, it gives an insight into the localization of the coherent
states viewed as wave packets on the line $\mathsf{x}\in \mathbb{R}$ and
their spreading in function of the rescaled ``time'' $\gamma$.

\begin{figure}[tbp]
\begin{center}
\includegraphics[width=9.5cm]{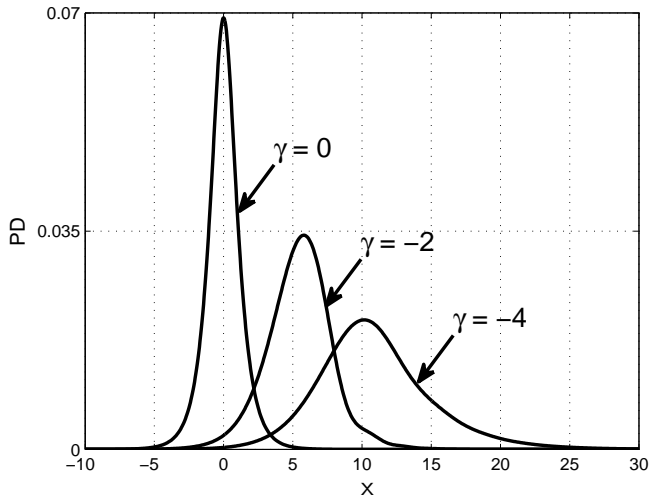}
\end{center}
\caption{Calculated probability densities (Eq. \protect\ref{locprobspat1})
on the real line $\mathsf{x}\in \mathbb{R}$ with $\mathfrak{J}=0$, $\protect%
\varsigma=\protect\epsilon/2=\protect\eta=1$, for the values $\protect\gamma %
=-4, \, -2,\, 0$. One can notice the spreading of the wave packet.}
\label{probdensfigpos}
\end{figure}

\section{Conclusion}

We have presented two methods for constructing families of coherent states
adapted to the quantum description of unbounded motions on the real line.

The first approach follows the Malkin-Manko treatment of quadratic
Hamiltonians and is more of algebraic nature, resting upon canonical
commutation rules and invariance principles. We have considered the example
of a particle submitted to a constant force (i.e. linear potential) and
obtained families of states fulfilling semi-classical exigences. We have
also given some insight about generalization to arbitrary potentials.

The second approach is of probabilistic nature. It provides a broad range of
possibilities in choosing the three main ingredients of the CS construction:
the function $E\mapsto \mathfrak{J}(E)$ on a classical level, and, on a
quantum level, the probability distributions $\mathfrak{J}\mapsto p_{%
\mathscr{E}}( \mathfrak{J})$ and the frequency function $\mathscr{E} \mapsto
\alpha_{\mathscr{E}}$. Of course, the selection should be ruled by the
requirement of manageable quantum operators combined with acceptable
semi-classical properties.

In a next publication we will examine in a more comprehensive way the
following points:

\begin{itemize}
\item[(i)] generalization of the first method to arbitrary potentials,

\item[(ii)] algebraic and domain properties of position and momentum
operators yielded by the second approach,

\item[(iii)] detailed comparison of the two approaches with regard to
localization properties in phase space and in configuration space.
\end{itemize}

\section*{Acknowledgement}

The work of VGB was partially supported by FAPESP, the Federal Targeted
Program "Scientific and scientific - pedagogical personnel of innovative
Russia", contract No P789 and Russia President grant SS - 1694.2012.2.
Gitman thank CNPq and FAPESP for permanent support.

\end{document}